\documentclass[floatfix,11pt,nofootinbib]{revtex4}
\usepackage{amssymb,amsmath}
\usepackage{graphicx,float}
\usepackage{indentfirst}

\newcommand{\beq}{\begin{equation}}
\newcommand{\eeq}{\end{equation}}
\newcommand{\bea}{\begin{eqnarray}}
\newcommand{\eea}{\end{eqnarray}}

\def\({\left(}
\def\){\right)}

\input{tcilatex}
\begin{document}

\title{Identifying metric spinor phase with axion field from axion-fermion
coupling and Weyl-Peccei Quinn spin transformations}
\author{Andr\'{e} Martorano Kuerten}
\email{martoranokuerten@hotmail.com}
\affiliation{Independent Researcher}

\begin{abstract}
Metric spinor phase of the Infeld-van der Waerden $\gamma $-formalism and
axion field were identified in Ref. \cite{amkafs}, by using Maxwell's
theory. Since axion couples with fermions, we will investigate Dirac's
theory to extend the work given in \cite{amkafs}, showing that is possible
again to identify this phase with the axion. By searching an exact
identification, we will work yet the spin transformations to adapt chiral
rotations and so to show that the metric spinor phase behaves exactly as
required for the axion field. We will study also the Dirac-Maxwell system
which provides a minimal Lagrangian approach for the axion classical sector
and, finally, we will obtain an explicit $2$-spinor description of magnetic
monopole and its charge.
\end{abstract}

\maketitle

\section{Introduction}

Axion theories have provided important results in several areas, such as 
\textrm{QCD}, \textrm{CP} symmetry problem, condensed matter and string
theory \cite{peccei,quinn,zhang,bakas}. The axion is also a good candidate
to explain cold dark matter \cite{preskill,abbott,fischler,gondolo}, even
though it is not directly observed in experiments. The axion/Maxwell
coupling is given by the following Lagrangian term%
\begin{equation}
\alpha F^{\mu \nu }F_{\mu \nu }^{\star }.
\end{equation}%
Since $A_{\mu }{}$ is the electromagnetic gauge potential and $e_{\mu \nu
\sigma \rho }$ are the Levi-Civita tensor components, the objects $\alpha $, 
$F_{\mu \nu }\doteqdot 2\nabla _{\lbrack \mu }A_{\nu ]}$ and $F_{\mu \nu
}^{\star }$ represent, respectively, the axion pseudo-scalar, Maxwell's
tensor and its Hodge dual defined by $F_{\mu \nu }^{\star }\doteqdot
(1/2)e_{\mu \nu }{}^{\sigma \rho }F_{\sigma \rho }$. In the usual cases the
term $F^{\mu \nu }F_{\mu \nu }^{\star }$ is taken to zero. In \cite%
{tiwari,tiwari2,tiwari3,tiwari4}, have been formulated a local dual
invariant electrodynamics (\textrm{LDIE}) formalism to add axionic fields in
electromagnetic equations.

On another hand, by using the Infeld-van der Waerden formalisms \cite%
{waerden,infeld2,infeld,corson,bade,cardoso,cardoso2,cardoso3}, geometric
sources for Infeld-van der Waerden electromagnetic fields have been defined
in \cite{kuerten}, such that in \cite{amkafs}, magnetic monopoles are
defined in a similar way. In this case, the spinor formulation of Maxwell
equations in $\gamma $-formalism yields an axion electrodynamics identical
with the found by Tiwari \cite{tiwari,tiwari2,tiwari3,tiwari4}. Based on
Weyl's representation, the Infeld-van der Waerden $\gamma \varepsilon $%
-formalisms have been useful for spinor formulation of General Relativity 
\cite{infeld}. Each $\gamma \varepsilon $-formalism is based on its spinor
metric: constant in the $\varepsilon $-formalism and depending locally on
the coordinates in the $\gamma $-formalism. In these formulations, the
spinor decomposition of $g_{\mu \nu }$ admits a local scale/phase freedom
given by $\varepsilon _{AB}\mapsto \left\vert \gamma \right\vert e^{\Theta
i}\varepsilon _{AB}$ and $\Sigma _{\mu }^{AA^{\prime }}\mapsto \left\vert
\gamma \right\vert ^{-1}\Sigma _{\mu }^{AA^{\prime }}$. Here, $\varepsilon
_{AB}$ is the metric spinor component\ and $\Sigma _{\mu }^{AA^{\prime }}$ a
connecting object component (or Infeld-van der Waerden symbol). Thus, the $%
\gamma $-formalism defines the metric spinor and the connecting objects as
follow%
\begin{equation}
\gamma _{AB}\doteqdot \left\vert \gamma \right\vert e^{\Theta i}\varepsilon
_{AB}\text{ \ \ and \ \ }\Upsilon _{\mu }^{AA^{\prime }}\doteqdot \left\vert
\gamma \right\vert ^{-1}\Sigma _{\mu }^{AA^{\prime }}.  \label{bd}
\end{equation}%
In \cite{amkafs}, by using electromagnetic theory, the identification $%
\alpha \sim \Theta $ has been established. It is known that classical
\textquotedblleft world\textquotedblright\ theories can be rewritten in a $2$%
-spinor version, since the linear group of unimodular complex $2\times 2$
matrices has a homomorphism two to one with the orthochronous proper Lorentz
group.

The Infeld-van Waerden formalisms was based on the homomorphism between the
Weyl and Lorentz groups \cite{afriat,amkafs}. Originally, this theory
provided a geometrical origin of the electromagnetic potential, since it
would lead to an imaginary part of the spinor connection trace in which
would satisfy the Weyl's principle of gauge invariance \cite{weyl}. Weyl
studied the relationship between the tetrad formalism for curved spacetime
and the parameter of the Dirac $4$-spinor phase transformation: if the
tetrad varies so the parameter varies too \cite{weyl,afriat}. Infeld and van
der Waerden considered Weyl's work to implement Dirac's theory in General
Relativity. However, this idea wasn't consolidated it should not be
understood on its original form \cite{honorinf}. On this interpretation, the
formalism would imply a relation between electric charge and spin, since the
scale/phase couples with each type of fermion. Unfortunately, the neutron
disabled this idea: as it has spin but no electric charge. Furthermore, the
interpretation of the imaginary part of the spinor connection trace impaired
some investigations of Maxwell's theory in the $\gamma $-formalism. In \cite%
{amkafs}, has been considered the electromagnetic potential as an external
physical entity. Thus, the physical significance of the phase and scale are
freedom to be reinterpred.

Since axion was identificate with Infeld-van der Waerden phase in \cite%
{amkafs}, by using electromagnetic fields, we will want to study outher
fields that interacts with $\alpha $. Fermion-axion coupling is given by the
following Lagrangian term \cite{raffelt}%
\begin{equation}
\widetilde{\Psi }\gamma ^{\mu }\Psi \partial _{\mu }\alpha ,  \label{af}
\end{equation}%
with $\Psi $ being the Dirac $4$-spinor, $\widetilde{\Psi }\doteqdot \Psi
^{\dagger }\gamma ^{0}$ its spinor adjoint and $\gamma ^{\mu }$ the Dirac
matrices. By wanting to repeat a similar result with the derived in \cite%
{amkafs}, we will investigate Dirac's theory in the $\gamma $-formalism.

We will use $\hbar =c=1$, as well as the signature $(+---)$. The index
symmetry/antisymmetry will be indicated by round/square brackets. The paper
will be organized as follows. In section \textrm{2} we will review the axion
electrodynamics provided by the $\gamma $-formalism. In the section \textrm{3%
}, we must obtain the axion-fermion coupling from $\gamma $-formalism and
establish the Weyl-Peccei Quinn transformations to identify the phase with
the axion, as well as, we will write Maxwell-Dirac theory to obtain an
explicit magnetic monopole $2$-spinor form and derive its effective magnetic
charge.

\section{Identifying Axion with Metric Spinor Phase: Maxwell Case}

In this section, we will review the work given in \cite{amkafs}. The axion
electrodynamics was originally postulated by Wilczek \cite{wilczek}. The
Lagrangian in which provides the axion electrodynamics can be represented by%
\begin{equation}
2F^{\mu \nu }F_{\mu \nu }+4\alpha F^{\mu \nu }F_{\mu \nu }^{\star }+A^{\mu
}j_{\mu },  \label{oae}
\end{equation}%
where $\mathcal{L}_{\text{M}}=2F^{\mu \nu }F_{\mu \nu }$ is the usual
Maxwell Lagrangian, $\mathcal{L}_{\text{CS}}=4\alpha F^{\mu \nu }F_{\mu \nu
}^{\star }$ the Chern-Simons term and $\mathcal{L}_{\text{I}}=A^{\mu }j_{\mu
}$ the interaction term with the electric current density $j_{\mu }$. In the
3-vector notation, we have $F^{\mu \nu }F_{\mu \nu }=\mathbf{B}^{2}-\mathbf{E%
}^{2}$ and $F^{\mu \nu }F_{\mu \nu }^{\star }=\mathbf{E}\bullet \mathbf{B}$. 
$\mathbf{E}$\ and $\mathbf{B}$\ are the electric and magnetic fields
respectively. The usual field equations for axion electrodynamics are given
by the expressions%
\begin{equation}
\partial ^{\mu }F_{\mu \nu }=4\pi j_{\nu }+\left( \partial ^{\mu }{}\alpha
\right) F_{\mu \nu }^{\star }\text{ \ \ and \ \ }\partial ^{\mu }F_{\mu \nu
}^{\star }=0.  \label{eom}
\end{equation}

In \cite{tiwari}, by studing a generic local dual electrodynamics, Tiwari
found the following axion electrodynamics equations adapt for us as 
\begin{equation}
\partial ^{\mu }F_{\mu \nu }=4\pi j_{\nu }+\left( \partial ^{\mu }{}\alpha
\right) F_{\mu \nu }^{\star }\text{ \ \ \ \ and \ \ \ \ }\partial ^{\mu
}F_{\mu \nu }^{\star }=4\pi m_{\nu }-\left( \partial ^{\mu }{}\alpha \right)
F_{\mu \nu }.  \label{sudtiw}
\end{equation}%
where $m_{\mu }$ is a magnetic current density. The first expression resumes
the effect caused from an axion field. The second can be understood as a
solution for the non observation of magnetic monopole in nature, since the
axion term can cancel its effects. In general, expressions (\ref{sudtiw})
are invariant when duality rotations%
\begin{equation}
\mathbb{F\mapsto UF}\text{ \ \ and \ \ }\mathbb{S\mapsto US}\text{, \ \ }%
\mathbb{F}=%
\begin{pmatrix}
F_{\mu \nu } \\ 
F_{\mu \nu }^{\star }%
\end{pmatrix}%
,\text{ \ }\mathbb{S}=%
\begin{pmatrix}
j_{\nu } \\ 
m_{\nu }%
\end{pmatrix}%
\text{ \ and }\mathbb{U}=%
\begin{pmatrix}
\cos \phi & \sin \phi \\ 
-\sin \phi & \cos \phi%
\end{pmatrix}%
.
\end{equation}%
are taken in account, with $\phi =\phi (x^{\alpha })$ and simultaneously if
the axion changes as $\partial _{\mu }\alpha \mapsto \partial _{\mu }\alpha
+\partial _{\mu }\phi $. By considering valid the Maxwell's equations, we
find from (\ref{sudtiw}) the equations 
\begin{equation}
\left( \partial ^{\mu }{}\alpha \right) F_{\mu \nu }^{\star }=0\text{ \ \ \
\ and \ \ \ \ }\left( \partial ^{\mu }{}\alpha \right) F_{\mu \nu }=4\pi
m_{\nu }.  \label{tiw}
\end{equation}

On another hand, the spacetime algebra in $\gamma $-formalism is represented
as follows%
\begin{equation}
g_{\mu \nu }=\Upsilon _{\mu }^{AA^{\prime }}\Upsilon _{\nu }^{BB^{\prime
}}\gamma _{AB}\gamma _{A^{\prime }B^{\prime }}.
\end{equation}%
The Einstein convention is adopted and each spinor index runs of $0$ to $1$ (%
$0^{\prime }$ to $1^{\prime }$). The object $\gamma _{AB}$ is the metric
spinor and $\Upsilon _{\mu }^{AA^{\prime }}$ the Infeld-van der Waerden
symbols \cite{infeld,cardoso,penrose}. Explicity, we have (\ref{bd}). $%
g_{\mu \nu }$ is the metric tensor component of a generic background.
Complex conjugation is denoted by $(\Upsilon _{\mu }^{AB^{\prime }})^{\ast
}=\Upsilon _{\mu }^{A^{\prime }B}$. Spinors and tensors are related by using
a hermitian matrix set $\Upsilon $, such as $v_{\mu }=\Upsilon _{\mu
}^{AA^{\prime }}v_{AA^{\prime }}$ and $v_{AA^{\prime }}=\Upsilon
_{AA^{\prime }}^{\mu }v_{\mu }$. We use the metric\ spinor to lower (or
raise) the spinor indexes: $\xi _{A}=\gamma _{BA}\xi ^{B}$, $\xi ^{A}=\gamma
^{AB}\xi _{B}$. The object $\gamma _{AB}$ is a skew-symmetric spinor
component. In the matrix form, we have%
\begin{equation}
(\gamma _{AB})=%
\begin{pmatrix}
0 & \gamma \\ 
-\gamma & 0%
\end{pmatrix}%
,\text{ \ \ with \ \ }\gamma \doteqdot \left\vert \gamma \right\vert
e^{\Theta i}.
\end{equation}%
in which $\left\vert \gamma \right\vert $ and $\Theta $ are real-valued
functions of $x^{\mu }$.

For generic spinors spinors $\xi ^{A}$ and $\zeta _{A}$, the covariant
derivative is given respectively as%
\begin{equation}
\nabla _{\mu }\xi ^{A}=\partial _{\mu }\xi ^{A}+\Xi _{\mu B}{}^{A}\xi ^{B}%
\text{ \ \ \ \ and \ \ \ \ }\nabla _{\mu }\zeta _{A}=\partial _{\mu }\zeta
_{A}-\Xi _{\mu A}{}^{B}\zeta _{B},
\end{equation}%
with $\Xi _{\mu B}{}^{A}$ a spinor connection. The complex component $\Xi
_{\mu A}{}^{A}$ can be written as \cite{cardoso}%
\begin{equation}
\Xi _{\mu A}{}^{A}=\partial _{\mu }\ln \left\vert \gamma \right\vert -2i\Xi
_{\mu },
\end{equation}%
with $\Xi \doteqdot -(1/2)\func{Im}\Xi _{\mu A}{}^{A}$. In $\gamma $%
-formalism, the compatibility metric $\nabla _{\alpha }g_{\mu \nu }=0$
yields the eigenvalue equations \cite{cardoso}%
\begin{equation}
\nabla _{\mu }\gamma _{AB}=i\beta _{\mu }\gamma _{AB}\text{ \ \ \ \ and \ \
\ \ }\nabla _{\mu }\gamma ^{AB}=-i\beta _{\mu }\gamma ^{AB}.  \label{eve}
\end{equation}%
with $\beta _{\mu }$ defined by%
\begin{equation}
\beta _{\mu }\doteqdot \partial _{\mu }\Theta +2\Xi _{\mu }.  \label{eve2}
\end{equation}%
As it is usual, the spinors $\xi ^{A}$ and $\zeta _{A}$ transform under the
action of the generalized Weyl gauge group, which can be expressed in the
component form as%
\begin{equation}
\Delta _{A}{}^{B}=\sqrt{\rho }e^{i\lambda }\delta _{A}{}^{B},  \label{gwt}
\end{equation}%
where $\rho >0$ is a real function and $\lambda $\ the gauge parameter of
the group. In flat spacetime, the choice $\Xi _{\mu A}{}^{B}=0$ can be taken
in account. Thus, $\Xi _{\mu }=0$ and $\left\vert \gamma \right\vert
=const>0 $, so that $\beta _{\mu }=\partial _{\mu }\Theta $. Here, we will
assume $\left\vert \gamma \right\vert =1$.

In this formalism, the spinor version of the Maxwell's tensor and its Hodge
are \cite{cardoso,penrose,carmeli}%
\begin{equation}
2F_{AA^{\prime }BB^{\prime }}=\gamma _{AB}f_{A^{\prime }B^{\prime }}+\gamma
_{A^{\prime }B^{\prime }}f_{AB},\text{ \ \ and \ \ }2F_{AA^{\prime
}BB^{\prime }}^{\star }=i\left( \gamma _{AB}f_{A^{\prime }B^{\prime
}}-\gamma _{A^{\prime }B^{\prime }}f_{AB}\right) ,  \label{mbs}
\end{equation}%
in which $f_{AB}=f_{(AB)}$ is called of Maxwell spinor. By defining the
complex tensor $F_{\mu \nu }^{(\pm )}\doteqdot F_{\mu \nu }\pm iF_{\mu \nu
}^{\star }$ and by taking the spinor forms (\ref{mbs}), we find%
\begin{equation}
F_{AA^{\prime }BB^{\prime }}^{(+)}=\gamma _{A^{\prime }B^{\prime }}f_{AB}%
\text{ \ \ \ \ and \ \ \ \ }F_{AA^{\prime }BB^{\prime }}^{(-)}=\gamma
_{AB}f_{A^{\prime }B^{\prime }}.  \label{+-}
\end{equation}%
The complex version of Maxwell equations $\nabla ^{\mu }F_{\mu \nu }^{(\pm
)}=4\pi j_{\nu }$ yield%
\begin{equation}
\nabla _{A^{\prime }}^{B}f_{AB}=2\pi j_{AA^{\prime }}+i\beta _{A^{\prime
}}^{B}f_{AB}.  \label{gee}
\end{equation}%
Since in the usual $\varepsilon $-formalism, the Maxwell equations with
magnetic monopoles $\nabla ^{\mu }F_{\mu \nu }^{(\pm )}=4\pi \left( j_{\nu
}\pm im_{\nu }\right) $ provides%
\begin{equation}
\nabla _{A^{\prime }}^{B}f_{AB}=2\pi \left( j_{AA^{\prime }}+im_{AA^{\prime
}}\right) ,  \label{me3}
\end{equation}%
follows the definition eleborate in \cite{amkafs}, i. e.,%
\begin{equation}
\beta _{A^{\prime }}^{B}f_{AB}\doteqdot 2\pi m_{AA^{\prime }}.  \label{ms}
\end{equation}%
With this definition, the expression (\ref{ms}) implies $\beta ^{\mu }F_{\mu
\nu }^{(\pm )}=4\pi m_{\nu }$. Since $\beta _{\mu }=\partial _{\mu }\Theta $
in flat spacetime, this equation becomes $\left( \partial ^{\mu }\Theta
\right) F_{\mu \nu }^{(\pm )}=4\pi m_{\nu }$. By expliciting $F_{\mu \nu
}^{(\pm )}$, we find%
\begin{equation}
\left( \partial ^{\mu }\Theta \right) F_{\mu \nu }=4\pi m_{\nu }\text{ \ \
and \ \ }\left( \partial ^{\mu }\Theta \right) F_{\mu \nu }^{\star }=0.
\label{amkt}
\end{equation}%
By comparing (\ref{amkt}) with (\ref{tiw}), the identification $\alpha \sim
\Theta $ is done.

On the Lagrangian viewpoint in Minkowski spacetime, the Maxwell Lagrangian
in $\gamma $-formalism is given by%
\begin{equation}
\mathcal{L}_{\text{M}}=2F_{\mu \nu }F^{\mu \nu }=\func{Re}\left[ \gamma
^{AC}\gamma ^{BD}f_{AB}f_{CD}\right] ,  \label{sml}
\end{equation}%
which the right side can be rewritten as follows%
\begin{equation}
\func{Re}\left[ \gamma ^{AC}\gamma ^{BD}f_{AB}f_{CD}\right] =\func{Re}\left[
\varepsilon ^{AC}\varepsilon ^{BD}f_{AB}f_{CD}\right] \cos \left( 2\Theta
\right) +\func{Im}\left[ \varepsilon ^{AC}\varepsilon ^{BD}f_{AB}f_{CD}%
\right] \sin \left( 2\Theta \right) .  \label{ek}
\end{equation}%
By taking the approximation $\gamma ^{AB}\simeq \varepsilon ^{AB}$ ($\Theta
\simeq 0$) and thus by considering%
\begin{equation}
\func{Re}\left[ \varepsilon ^{AC}\varepsilon ^{BD}f_{AB}f_{CD}\right] \simeq
2F^{\mu \nu }F_{\mu \nu }\text{ \ \ and \ \ }\func{Im}\left[ \varepsilon
^{AC}\varepsilon ^{BD}f_{AB}f_{CD}\right] \simeq 2F^{\mu \nu }F_{\mu \nu
}^{\star },  \label{app}
\end{equation}%
we find from (\ref{sml}), the theory%
\begin{equation}
\mathcal{L}_{\text{M}}\simeq 2F^{\mu \nu }F_{\mu \nu }+4\Theta F^{\mu \nu
}F_{\mu \nu }^{\star }.  \label{aec}
\end{equation}%
Thus, when $\Theta \simeq 0$, the invariant form $\func{Re}\left[
f^{AB}f_{AB}\right] $ in $\gamma $-formalism generates the usual Maxwell
Lagrangian plus the axionic Chern-Simons term $\mathcal{L}_{\text{CS}%
}\doteqdot \Theta F^{\mu \nu }F_{\mu \nu }^{\star }$ in which relates the
axion like Infeld-van der Waerden phase-electromagnetic coupling.

\section{Phase-fermion Coupling and Weyl-PQ transformations}

We will follow Ref. \cite{cardoso3} to present Dirac's theory. Outher works
about Dirac's theory in the Infeld-van der Waerden formalisms are found in 
\cite{infeld,bade,dcardoso}. In $2$-component spinor formalism, Dirac
equations in relativistic spacetimes can be stated as follow%
\begin{equation}
i\nabla _{AA^{\prime }}\psi ^{A}=\mu \chi _{A^{\prime }}\text{ \ \ and\ \ \ }%
i\nabla ^{AA^{\prime }}\chi _{A^{\prime }}=\mu \psi ^{A}.  \label{sde}
\end{equation}%
$\psi ^{A}$ and $\chi _{A^{\prime }}$ are, respectively, right handed and
left handed $2$-spinors. $\mu \doteqdot -m/\sqrt{2}$, where minus sign is
placed according with our purpose. Thanks to the fact that in the $\gamma $%
-formalism we have eigenvalue equations for $\gamma _{AB}$, (\ref{sde}) is
equivalent to%
\begin{equation}
\nabla ^{AA^{\prime }}\psi _{A}-i\beta ^{AA^{\prime }}\psi _{A}=i\mu \chi
^{A^{\prime }}\text{ \ \ and \ \ }\nabla _{AA^{\prime }}\chi ^{A^{\prime
}}-i\beta _{AA^{\prime }}\chi ^{A^{\prime }}=i\mu \psi _{A}.  \label{sde2}
\end{equation}%
The $\varepsilon $-formalism version of (\ref{sde2}) is obtained by taking $%
\beta _{AA^{\prime }}$ to zero. Dirac's fields which satisfy respectively (%
\ref{sde}) and (\ref{sde2}) are given by the systems%
\begin{equation}
\overline{\text{D}}=\left\{ \left( \psi ^{A},\chi _{A^{\prime }}\right)
,\left( \chi _{A},\psi ^{A^{\prime }}\right) \right\} \text{ \ \ and \ \ }%
\underline{\text{D}}=\left\{ \left( \psi _{A},\chi ^{A^{\prime }}\right)
,\left( \chi ^{A},\psi _{A^{\prime }}\right) \right\} .
\end{equation}%
By using metric spinors depending on the formalism considered, $\underline{%
\text{D}}$ is obtained from $\overline{\text{D}}$. We stress that only $%
\underline{\text{D}}$ couples with $\beta $-terms.

If we want to obtain usual covariant Dirac's equation in the $\varepsilon $%
-formalism, we must define the $4$-component Dirac's field $\Psi $ as follows%
\begin{equation}
\Psi \doteqdot 
\begin{pmatrix}
\psi _{A} \\ 
\chi ^{A^{\prime }}%
\end{pmatrix}%
.  \label{4s}
\end{equation}%
This choice is valid, since we have $\Psi \mapsto e^{i\lambda }\Psi $ under
the original Weyl's group action: $\psi _{A}\mapsto e^{i\lambda }\psi _{A}$
and $\chi ^{A^{\prime }}\mapsto e^{i\lambda }\chi ^{A^{\prime }}$. Original
Weyl's group is recovered by taking $\rho =1$ in (\ref{gwt}). If our
interest concerns only on axion/fermion coupling, we can work in flat
spacetime. In this background and in the $\varepsilon $-formalism, equation (%
\ref{sde2}) becomes%
\begin{equation}
\partial ^{AA^{\prime }}\psi _{A}=i\mu \chi ^{A^{\prime }}\text{ \ \ and \ \ 
}\partial _{AA^{\prime }}\chi ^{A^{\prime }}=i\mu \psi _{A}.  \label{msde}
\end{equation}%
On the $\varepsilon $-formalism, covariant derivative $\nabla _{AA^{\prime
}} $ is taken by using Infeld-van der Waerden symbols: $\nabla _{AA^{\prime
}}=\Sigma _{AA^{\prime }}^{\mu }\nabla _{\mu }$ and $\nabla ^{AA^{\prime
}}=\varepsilon ^{AB}\varepsilon ^{A^{\prime }B^{\prime }}\Sigma _{BB^{\prime
}}^{\mu }\nabla _{\mu }$. In Minkowski universe $\sqrt{2}\Sigma _{AA^{\prime
}}^{\mu }=\sigma _{AA^{\prime }}^{\mu }$ as well as $\nabla _{\mu }=\partial
_{\mu }$. $\left( \sigma _{AA^{\prime }}^{\mu }\right) =\left( \mathbb{I}%
,\sigma _{AA^{\prime }}^{i}\right) $ where $\left( \sigma _{AA^{\prime
}}^{i}\right) $ are the Pauli matrices and $\mathbb{I}$ the $2\times 2$
unity matrix. If taken in account the Weyl's representation, the Dirac's
matrices become%
\begin{equation}
\gamma ^{0}=%
\begin{pmatrix}
\mathbb{O} & \sigma ^{0} \\ 
\sigma ^{0} & \mathbb{O}%
\end{pmatrix}%
\text{ \ \ and \ \ }\gamma ^{i}=%
\begin{pmatrix}
\mathbb{O} & \sigma ^{i} \\ 
-\sigma ^{i} & \mathbb{O}%
\end{pmatrix}%
.
\end{equation}

If we use the definition (\ref{4s}) and Dirac's matrices in the Weyl's
representation, Dirac's equation is obtained from (\ref{msde}). In fact, we
have%
\begin{equation}
\begin{pmatrix}
\mathbb{O} & \frac{1}{\sqrt{2}}\sigma _{AA^{\prime }}^{\mu }\partial _{\mu }
\\ 
\frac{1}{\sqrt{2}}\sigma _{\mu }^{AA^{\prime }}\partial ^{\mu } & \mathbb{O}%
\end{pmatrix}%
\begin{pmatrix}
\psi _{A} \\ 
\chi ^{A^{\prime }}%
\end{pmatrix}%
=\frac{1}{\sqrt{2}}\gamma ^{\mu }\partial _{\mu }\Psi \text{ \ \ and \ \ }%
\mu 
\begin{pmatrix}
\psi _{A} \\ 
\chi ^{A^{\prime }}%
\end{pmatrix}%
=-\frac{m}{\sqrt{2}}\Psi ,
\end{equation}%
so that we obtain $\left( i\gamma ^{\mu }\partial _{\mu }-m\right) \Psi =0$,
which is the familiar covariant Dirac's equation.

By considering the Euler-Lagrange equations, (\ref{msde}) are derived from
the Lagrangian%
\begin{equation}
\mathcal{L}_{\text{D}}\left[ \overline{\text{D}},\partial \overline{\text{D}}%
\right] =i\psi ^{A^{\prime }}\partial _{A^{\prime }A}\psi ^{A}+i\chi
_{A}\partial ^{AA^{\prime }}\chi _{A^{\prime }}-\mu \left( \psi ^{A}\chi
_{A}+\psi ^{A^{\prime }}\chi _{A^{\prime }}\right) ,  \label{le}
\end{equation}%
since the relationship%
\begin{equation}
\psi ^{A}\chi _{A}+\psi ^{A^{\prime }}\chi _{A^{\prime }}=-\psi _{A}\chi
^{A}-\psi _{A^{\prime }}\chi ^{A^{\prime }}=\widetilde{\Psi }\Psi ,
\label{a1}
\end{equation}%
is satisfied in both formalisms and%
\begin{equation}
i\psi ^{A^{\prime }}\partial _{A^{\prime }A}\psi ^{A}+i\chi _{A}\partial
^{AA^{\prime }}\chi _{A^{\prime }}=i\psi _{A^{\prime }}\partial ^{AA^{\prime
}}\psi _{A}+i\chi ^{A}\partial _{AA^{\prime }}\chi ^{A^{\prime }}=\frac{1}{%
\sqrt{2}}i\widetilde{\Psi }\gamma ^{\mu }\partial _{\mu }\Psi ,  \label{a}
\end{equation}%
only in the $\varepsilon $-formalism. Notation given in (\ref{le}) denotes
that $\mathcal{L}_{\text{D}}$ depends of $\overline{\text{D}}$ and
derivatives. In $2$-spinor notation, we have yet $\widetilde{\Psi }\doteqdot
\Psi ^{\dagger }\gamma ^{0}=%
\begin{pmatrix}
\chi ^{A} & \psi _{A^{\prime }}%
\end{pmatrix}%
$. If $\rho =1$ in (\ref{gwt}), we note that $\widetilde{\Psi }$ transforms
as $\widetilde{\Psi }\mapsto e^{-i\lambda }\widetilde{\Psi }$, which must be
in this specific case. Thus, if we use (\ref{a1}) and (\ref{a}), the
Lagrangian (\ref{le}) assumes the form $\mathcal{L}_{\text{D}}\left[ 
\overline{\text{D}},\partial \overline{\text{D}}\right] =\mathcal{L}_{\text{D%
}}\left[ \underline{\text{D}}\text{,}\partial \underline{\text{D}}\right] =i%
\widetilde{\Psi }\gamma ^{\mu }\partial _{\mu }\Psi -m\widetilde{\Psi }\Psi $%
, which is the\ usual Dirac's Lagrangian. The factor $\sqrt{2}$ is absorbed
by the redefinition $\mathcal{L}_{\text{D}}\mapsto \sqrt{2}\mathcal{L}_{%
\text{D}}$.

Since that in the $\varepsilon $-formalism, Dirac's theory is obtained from (%
\ref{le}), we will consider its index configuration as starting form in the $%
\gamma $-formalism. In this formalism, equations (\ref{sde2}) in flat
spacetime are rewritten by putting $\beta _{AA^{\prime }}=\partial
_{AA^{\prime }}\Theta $. The equations (\ref{sde}) have the same form that
those given in (\ref{msde}), in both formalisms. Let us study (\ref{le}).
Equation (\ref{a1}) is valid also in the $\gamma $-formalism. However, a
change in the index configuration of (\ref{a}) yields, in the $\gamma $%
-formalism, the expression%
\begin{equation}
i\psi ^{A^{\prime }}\partial _{A^{\prime }A}\psi ^{A}+i\chi _{A}\partial
^{AA^{\prime }}\chi _{A^{\prime }}=i\psi _{A^{\prime }}\partial ^{AA^{\prime
}}\psi _{A}+i\chi ^{A}\partial _{AA^{\prime }}\chi ^{A^{\prime }}+\psi
_{A^{\prime }}\psi _{A}\partial ^{AA^{\prime }}\Theta +\chi ^{A}\chi
^{A^{\prime }}\partial _{AA^{\prime }}\Theta ,
\end{equation}%
due to eigenvalue equations. Since that%
\begin{equation}
\psi _{A^{\prime }}\psi _{A}\partial ^{AA^{\prime }}\Theta +\chi ^{A}\chi
^{A^{\prime }}\partial _{AA^{\prime }}\Theta =\frac{1}{\sqrt{2}}\widetilde{%
\Psi }\gamma ^{\mu }\Psi \partial _{\mu }\Theta ,  \label{pfc}
\end{equation}%
we obtain (\ref{af}) from (\ref{pfc}), by simply indentifing $\alpha \sim
\Theta $. Thus, in the $\gamma $-formalism, we have the functional relation%
\begin{equation}
\mathcal{L}_{\text{D}}\left[ \overline{\text{D}},\partial \overline{\text{D}}%
\right] =\mathcal{L}_{\text{D}}\left[ \underline{\text{D}}\text{,}\partial 
\underline{\text{D}},\partial \Theta \right] =i\widetilde{\Psi }\gamma ^{\mu
}\partial _{\mu }\Psi -m\widetilde{\Psi }\Psi +\widetilde{\Psi }\gamma ^{\mu
}\Psi \partial _{\mu }\Theta ,  \label{da1}
\end{equation}%
which represents an axion-like coupling between Dirac fields and $\Theta $.
Again, $\mathcal{L}_{\text{D}}$\ absorbed $\sqrt{2}$. Therefore, we have
identified the Infeld-van der Waerden phase with axion field.

\subsection{Weyl-PQ (Peccei-Quinn) Spin Transformations}

We have identified the metric spinor phase with the axion field from Maxwell
and Dirac theories. Moreover, such identifications are \textquotedblleft
poors\textquotedblright\ since the behavior of $\Theta $ on chiral rotation
has not been established. By considering (\ref{gwt}), we have seen that the
Dirac's spinor is defined in terms of Weyl's $2$-spinors according with the
action of the Weyl group. On another hand, we have also the chiral
rotations: $\Psi \mapsto \widetilde{\Psi }=e^{i\zeta \gamma _{5}}\Psi $. The
original Peccei-Quinn procedure \cite{peccei} states that, on a chiral (PQ)
transformation, an axion-like pseudo-scalar $\theta $ transforms as%
\begin{equation}
\theta \mapsto \widetilde{\theta }=\theta -2\zeta .  \label{pst}
\end{equation}%
The transformation (\ref{pst}) is an elegant solution for solve the \textrm{%
CP}-problem. Here, our strategy is identify $\Theta $ with $\alpha $ by
using a similar way with the developed by Infeld and van der Waerden. The
key idea of the Infeld-van der Waerden unification is to consider the gauge
behavior of the electromagnetic potential $A_{\mu }$, i. e.,%
\begin{equation}
A_{\mu }\mapsto \widehat{A}_{\mu }=A_{\mu }-\partial _{\mu }\lambda .
\label{pst3}
\end{equation}%
Since the object $\Xi _{\mu }$ transforms as (see for example \cite{cardoso})%
\begin{equation}
\widehat{\Xi }_{\mu }=\Xi _{\mu }-\frac{1}{2}\func{Im}\left( \partial _{\mu
}\ln \Delta \right) ,  \label{ft}
\end{equation}%
by putting $\Delta \doteqdot \det (\Delta _{A}{}^{B})=e^{2i\lambda }$ in (%
\ref{ft}), we find%
\begin{equation}
\Xi _{\mu }\mapsto \widehat{\Xi }_{\mu }=\Xi _{\mu }-\partial _{\mu }\lambda
.  \label{pst2}
\end{equation}%
By looking (\ref{pst3}) and (\ref{pst2}), it is suggested the identification 
$\Xi _{\mu }\sim A_{\mu }$.

Our strategy is verify if $\Theta $ behaves as (\ref{pst}) when a PQ
transformation is taken in account. A similar problem was solved in \cite%
{toyoda}, where the author worked $\Xi _{\mu }$ as being a mixture of polar
and axial vectors. Our first step is to find spin transformations which
represent simultaneously Weyl and PQ rotations. In 2-spinor language, $\Psi
\mapsto e^{i\zeta \gamma _{5}}\Psi $ is translated as $\widetilde{\psi }%
_{A}=e^{i\zeta }\psi _{A}$ and $\widetilde{\chi }^{A^{\prime }}=e^{-i\zeta
}\chi ^{A^{\prime }}$. Thus, we need to obtain the spin transformations
where, in general, $\overline{\psi }_{A}=e^{i\left( \lambda +\zeta \right) }$%
\ and $\overline{\chi }^{A^{\prime }}=e^{i\left( \lambda -\zeta \right)
}\chi ^{A^{\prime }}$\ are verified. The notation ($\overline{n}$) denotes a
composite Weyl ($\widehat{n}$)-PQ ($\widetilde{n}$) transformation.

Let us consider the general spin transformation of the metric spinor:%
\begin{equation}
\overline{\gamma }_{AB}=\Delta _{A}{}^{C}\Delta _{B}{}^{D}\gamma _{CD}.
\label{smt}
\end{equation}%
Now, we will consider the composition%
\begin{equation}
\Delta _{A}{}^{B}=\Delta _{A}^{\circ }{}^{C}\Delta _{C}^{\bullet }{}^{B}=%
\sqrt{\Delta ^{\circ }\Delta ^{\bullet }}\delta _{A}{}^{B},
\end{equation}%
where $\Delta ^{\circ }$ and $\Delta ^{\bullet }$ are the determinants of
the Weyl $(\Delta _{A}^{\circ }{}^{B})$ and PQ $(\Delta _{A}^{\bullet
}{}^{B})$ rotations. For a Weyl transformation we have $\widehat{\gamma }%
_{AB}=\Delta ^{\circ }\gamma _{AB}$ while for PQ transformation we will
supose $\widetilde{\gamma }_{AB}=\Delta ^{\bullet }(\widetilde{\gamma }%
\varepsilon _{AB})$. With such considerations, (\ref{smt}) yields%
\begin{equation}
\overline{\gamma }_{AB}=\Delta ^{\circ }\Delta ^{\bullet }(\widetilde{\gamma 
}\varepsilon _{AB}).
\end{equation}%
As we will verify later, a PQ transformation can be implemented by requiring%
\begin{equation}
\overline{\gamma }_{AB}=\Delta ^{\circ }\gamma _{AB}\text{ \ \ }%
\Leftrightarrow \text{ \ \ }\widetilde{\gamma }=(\Delta ^{\bullet
})^{-1}\gamma .  \label{pt}
\end{equation}%
From (\ref{pt}), we find $\overline{\gamma }_{AB}=\widehat{\gamma }_{AB}$.
By implementing $\overline{\gamma }^{AB}\overline{\gamma }_{AB}=\gamma
^{AB}\gamma _{AB}$, we deduce%
\begin{equation}
\overline{\gamma }^{AB}=\left( \Delta ^{\circ }\right) ^{-1}\gamma ^{AB}%
\text{ \ \ and \ \ }\widetilde{\gamma }^{-1}=\Delta ^{\bullet }\gamma ^{-1}.
\end{equation}

The transformation of generic spinors $v_{A}$ and $u^{A}=\gamma ^{AB}u_{B}$
are then given by%
\begin{equation}
\overline{v}_{A}=\sqrt{\Delta ^{\circ }\Delta ^{\bullet }}v_{A}\text{ \ \
and \ \ }\overline{u}^{A}=\sqrt{\left( \Delta ^{\circ }\right) ^{-1}\Delta
^{\bullet }}u^{A}.
\end{equation}%
Each separated rotation acts as%
\begin{equation}
\widehat{v}_{A}=\sqrt{\Delta ^{\circ }}v_{A},\text{ \ \ }\widehat{u}^{A}=%
\sqrt{\left( \Delta ^{\circ }\right) ^{-1}}u^{A},\text{ \ \ }\widetilde{v}%
_{A}=\sqrt{\Delta ^{\bullet }}v_{A}\text{ \ \ and \ \ }\widetilde{u}^{A}=%
\sqrt{\Delta ^{\bullet }}u^{A}.
\end{equation}%
Since $\Delta ^{\circ }=e^{2\lambda i}$ and $\Delta ^{\bullet }=e^{2\zeta i}$%
, the $2$-component fermions transform as%
\begin{equation}
\overline{\psi }_{A}=e^{i\left( \lambda +\zeta \right) }\psi _{A}\text{ \ \
and \ \ }\overline{\chi }^{A^{\prime }}=e^{i\left( \lambda -\zeta \right)
}\chi ^{A^{\prime }},  \label{gt}
\end{equation}%
or, particularly, $\widehat{\psi }_{A}=e^{i\lambda }\psi _{A}$ ($\widehat{%
\chi }^{A^{\prime }}=e^{i\lambda }\chi ^{A^{\prime }}$) (Weyl) and $%
\widetilde{\psi }_{A}=e^{i\zeta }\psi _{A}$ ($\widetilde{\chi }^{A^{\prime
}}=e^{-i\zeta }\chi ^{A^{\prime }}$) (PQ).

As we have seen, the Weyl-PQ transformation is possible if $\widetilde{%
\gamma }=(1/\Delta ^{\bullet })\gamma $ is satisfied. In Minkowski spacetime
we have $\gamma =e^{i\Theta }$, such that from a PQ rotation, we obtain $e^{i%
\widetilde{\Theta }}=(e^{i\Theta }/\Delta ^{\bullet })$. Therefore $%
\widetilde{\Theta }=\Theta +i\ln \Delta ^{\bullet }+2n\pi $, which implies%
\begin{equation}
\Theta \mapsto \widetilde{\Theta }=\underset{\text{PQ-behavior}}{\underbrace{%
\Theta -2\zeta }}+2n\pi ,\text{ \ \ }n\in \mathbb{Z}.
\end{equation}%
since $\Delta ^{\bullet }=e^{2\zeta i}$. Thus, we have demonstrated that $%
\Theta $ satisfies the PQ\ requeriment (\ref{pst}).

We must note that the invariant form $\overline{u}^{A}\overline{v}_{A}$ is
broken in our formulation: $\overline{u}^{A}\overline{v}_{A}=e^{2\zeta
i}u^{A}v_{A}$. This property is a formal declaration of no chiral symmetry
for massive fermion terms. The invariant indexes configurations are now
given by%
\begin{equation*}
_{AA^{\prime }},\text{ \ \ }^{AA^{\prime }},\text{ \ \ }_{AA^{\prime
}}{}^{AA^{\prime }},\text{ \ \ }_{AA^{\prime }BB^{\prime }}{}^{AA^{\prime
}BB^{\prime }},...,.
\end{equation*}

\subsection{Maxwell-Dirac system and magnetic monopoles}

When coupled with Maxwell fields, the Dirac equations in curved spacetime
are taken by putting $\nabla _{AA^{\prime }}\mapsto \nabla _{AA^{\prime
}}-ieA_{AA^{\prime }}$ in (\ref{sde2}), i. e.,%
\begin{equation}
\left( \nabla ^{AA^{\prime }}-ieA^{AA^{\prime }}-i\beta ^{AA^{\prime
}}\right) \psi _{A}=i\mu \chi ^{A^{\prime }}\text{ \ \ and \ \ }\left(
\nabla _{AA^{\prime }}-ieA_{AA^{\prime }}-i\beta _{AA^{\prime }}\right) \chi
^{A^{\prime }}=i\mu \psi _{A},  \label{dam}
\end{equation}%
with $A_{\mu }$\ being an electromagnetic potential component. Since $%
f_{AB}=\nabla _{(A}^{A^{\prime }}A_{B)A^{\prime }}$ in the spinor language,
equations (\ref{gee}) can be rewritten as follows%
\begin{equation}
\left( \nabla _{A^{\prime }}^{B}-i\beta _{A^{\prime }}^{B}\right) \nabla
_{A}^{B^{\prime }}A_{BB^{\prime }}=e\left( \psi _{A}\psi _{A^{\prime }}+\chi
_{A}\chi _{A^{\prime }}\right) .  \label{gee2}
\end{equation}%
Here, we have used for the electric current density the expression \cite%
{penrose}%
\begin{equation}
j_{AA^{\prime }}=e\left( \psi _{A}\psi _{A^{\prime }}+\chi _{A}\chi
_{A^{\prime }}\right) ,  \label{sedc}
\end{equation}%
where $e$\ is the electric charge.

Let us consider the equations (\ref{amkt}) in curved spacetime:%
\begin{equation}
\nabla ^{\mu }F_{\mu \nu }^{(\pm )}=4\pi j_{\nu }\text{ \ \ and \ \ }\beta
^{\mu }F_{\mu \nu }^{(\pm )}=m_{\nu }.
\end{equation}%
Since $\beta ^{\lbrack \mu }\beta ^{\nu ]}=0$, the expression $\beta ^{\mu
}F_{\mu \nu }^{(\pm )}$ provides $\beta ^{\mu }m_{\mu }=0$. By applying the
covariant derivative in $\beta ^{\mu }F_{\mu \nu }^{(\pm )}=4\pi m_{\nu }$,
we obtain%
\begin{equation}
W^{\mu \nu }F_{\mu \nu }+8\pi \left( \beta ^{\mu }j_{\mu }+\nabla ^{\mu
}m_{\mu }\right) =0\text{ \ \ \ \ and \ \ \ }W^{\mu \nu }F_{\mu \nu }^{\star
}=0,  \label{ivwm}
\end{equation}%
where $W_{\mu \nu }\doteqdot 2\partial _{\lbrack \mu }\Xi _{\nu ]}$ is the
Infeld-van der Waerden curvature bivector. By remembering that $\Xi _{\mu
}=-(1/2)\func{Im}\Xi _{\mu A}{}^{A}$,\ if the spacetime is flat, we obtain $%
\partial ^{\mu }m_{\mu }=-j_{\mu }\partial ^{\mu }\Theta $ such that 
\begin{equation}
m_{\mu }=-\left( \Theta -\mathcal{C}\right) j_{\mu },  \label{mmc2}
\end{equation}%
where $\mathcal{C}$ is a constant and we have used yet the null divergence
for electric sources: $\partial ^{\mu }j_{\mu }=0$. Since in this case $%
\beta ^{\mu }m_{\mu }=-\Theta j_{\mu }\partial ^{\mu }\Theta =0$, we will
obtain $2\Theta \partial ^{\mu }\left( \Theta j_{\mu }\right) =\Theta
\partial ^{\mu }\left( \Theta j_{\mu }\right) $ implying $\partial ^{\mu
}m_{\mu }=0$ and $\beta ^{\mu }j_{\mu }=0$. Thus, in flat background, we
have the divergence/orthogonality relationships%
\begin{equation}
\partial ^{\mu }j_{\mu }=0=\partial ^{\mu }m_{\mu }\text{ \ \ and \ \ }%
j_{\mu }\partial ^{\mu }\Theta =0=m_{\mu }\partial ^{\mu }\Theta .
\end{equation}%
By taking (\ref{sedc}), the expression (\ref{mmc2}) in the $2$-spinor form
is so rewritten as%
\begin{equation}
m_{AA^{\prime }}=g_{eff}\left( \psi _{A}\psi _{A^{\prime }}+\chi _{A}\chi
_{A^{\prime }}\right) ,\text{ \ \ with \ \ }g_{eff}\doteqdot -e\left( \Theta
-\mathcal{C}\right) .  \label{2pmm}
\end{equation}%
The term $g_{eff}$ in (\ref{2pmm}) can be understood as an effective
magnetic charge. From a PQ transformation, the effective magnetic charge
behaves 
\begin{equation}
g_{eff}\mapsto \widetilde{g}_{eff}=g_{eff}+2e\left( \zeta -n\pi \right) ,
\end{equation}%
such that $\widetilde{g}_{eff}=g_{eff}$ when $\zeta =n\pi $, with $n\in 
\mathbb{Z}$.

\section{Conclusion}

We have obtained an axion-fermion coupling from a minimal Lagrangian
formulation, since the axion-Dirac Lagrangian density $\mathcal{L}_{\text{D}}%
\left[ \underline{\text{D}}\text{,}\partial \underline{\text{D}},\partial
\alpha \right] $ can be derived from $\mathcal{L}_{\text{D}}\left[ \overline{%
\text{D}},\partial \overline{\text{D}}\right] $ if $\alpha \sim \Theta $. As
also we recapitulate, the Maxwell Lagrangian density $\mathcal{L}_{\text{M}}$
provides an electromagnetic-axion coupling if $\alpha \sim \Theta \simeq 0$: 
$\mathcal{L}_{\text{M}}\simeq 2F^{\mu \nu }F_{\mu \nu }+4\mathcal{L}_{\text{%
CS}}$. Thus, we can derive of the $\gamma $-formalism an axionic classical
sector from the compact form:%
\begin{equation*}
\mathcal{L}_{\text{D}}\left[ \overline{\text{D}},\partial \overline{\text{D}}%
\right] +\mathcal{L}_{\text{M}}.
\end{equation*}%
Thus, the scientific use of our work can be justified by reductionist
arguments. However, the behavior of $\Theta $ by PQ rotations was not clear.
Thus, we have elaborate a Weyl-PQ approach in which the PQ rotations are
implemented in the usual spin transformations, providing a Weyl-PQ spin
unified formalism. By using this new formulation, we have shown that the
spacetime phase behaves exactly as the axion is supposed to transform when
the PQ group acts. Thus, we can conclude that the metric spinor phase $%
\Theta $ behaves geometrically as an axion field, as well as, it generates
axion-like couplings with Maxwell and Dirac fields. Finally, we obtained an
explicit $2$-spinor structure for magnetic monolope source such as its
effective charge. A rigorous formulation of the Weyl-PQ transformations in
the Infeld-van der Waerden formalisms and a more accurate study of the
Dirac-Maxwell system must be worked in a future paper.

\section*{Acknowledgments}

The author would like to thank A. Fernandes-Silva for discussions.

\end{document}